\documentclass[pre,twocolumn,showpacs,prl,preprintnumbers,amsmath,amssymb]{revtex4}
\usepackage{graphicx,epsfig}% Include figure files
\usepackage{dcolumn}% Align table columns on decimal point
\usepackage{bm}% bold math
\usepackage{psfrag}
\usepackage{plain}
\usepackage{tabularx}
\usepackage[latin1]{inputenc}
\usepackage{amsmath}
\begin{document}
\draft

\title{Saltation transport on Mars}
\author{Eric J. R. Parteli$^1$ and Hans J. Herrmann$^{2,3}$}
\affiliation{1. Institut f\"ur Computerphysik, ICP, Universit\"at Stuttgart, Pfaffenwaldring 27, 70569 Stuttgart, Germany. \\ 2. Computational Physics, IfB, ETH H\"onggerberg, HIF E 12, CH-8093, Z\"urich, Switzerland. \\ 3. Departamento de F\'{\i}sica, Universidade Federal do Cear\'a - 60455-760, Fortaleza, CE, Brazil.}

\date{\today}

\begin{abstract}
We present the first calculation of saltation transport and dune formation on Mars and compare it to real dunes. We find that the rate at which grains are entrained into saltation on Mars is one order of magnitude higher than on Earth. With this fundamental novel ingredient, we reproduce the size and different shapes of Mars dunes, and give an estimate for the wind velocity on Mars.

\end{abstract}


\maketitle

The surprising discovery of sand dunes on Mars by Mariner 9 in 1971 has challenged geologists and planetary physicists over years with the following enigma: Could Mars dunes have been formed by today's martian thin atmosphere? Bagnold \cite{Bagnold_1941} first noticed that the sand of aeolian dunes is transported through {\em{saltation}}, which consists of grains travelling in a sequence of ballistic trajectories and producing a {\em{splash}} of new ejected grains when colliding back onto the soil. But saltation occurs only if the wind friction speed $u_{\ast}$ --- which, together with the air density ${\rho}_{\mathrm{fluid}}$, is used to define the air shear stress $\tau = {\rho}_{\mathrm{fluid}}u_{\ast}^2$ --- is larger than a minimal threshold $u_{{\ast}{\mathrm{t}}}$. Because the atmospheric density of Mars is almost 100 times lower than on earth, only winds of strength one order of magnitude higher than that on our planet are capable of mobilizing the basaltic grains of mean diameter $d=500$ ${\mu}$m \cite{Edgett_and_Christensen_1991} which constitute Mars dunes \cite{Greeley_et_al_1980,Iversen_and_White_1982}. Such winds occasionally can occur on Mars \cite{Sutton_et_al_1978,Moore_1985,Sullivan_et_al_2005}. Here we find that the strong erosion due to these winds allows to understand dune formation on the red planet under present conditions. 

Once saltation starts, the number of saltating grains first increases exponentially due to the multiplicative process inherent in the splash events. But since the wind loses momentum to accelerate the grains, the flux {\em{saturates}} after a transient distance \cite{Owen_1964,Anderson_and_Haff_1988,McEwan_and_Willetts_1991,Butterfield_1993,Andreotti_2004}. At saturation, the air shear stress ${\tau}_{\mathrm{a}}$ decreases to the threshold ${\tau}_{{\mathrm{t}}} = {\rho}_{\mathrm{fluid}}u_{{\ast}{\mathrm{t}}}^2$, while the flux of grains increases to its maximum value. This transient phenomenon introduces the only relevant length scale in the physics of dunes: Any sand patch which is shorter than this saturation distance will be eroded and disappear. On the other hand, large enough sand hills develop a slip face wherever the slope exceeds the angle of repose, ${\theta}_{\mathrm{r}} = 34^{\circ}$ \cite{Bagnold_1941}. 

\begin{figure}
  \begin{center}
   \includegraphics[width=1.0 \columnwidth]{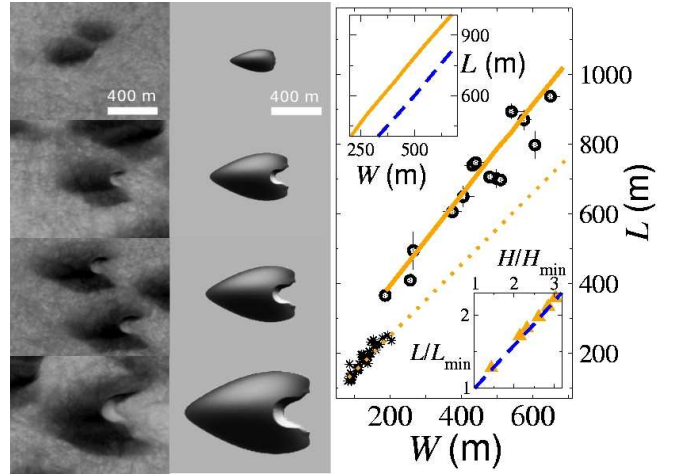}	
 \caption{Barchans in the Arkhangelsky crater, $41.0^{\circ}$S, $25.0^{\circ}$W on Mars: Mars Orbiter Camera (MOC) images on the left (image courtesy of NASA$/$JPL$/$MSSS) and calculated dunes on the right. Main plot: $L$ vs $W$ of the Arkhangelsky (circles) and north polar barchans at $77.6^{\circ}$N, $103.6^{\circ}$W (stars). Calculations of Arkhangelsky (north polar) dunes are represented by the continuous (dotted) line, obtained with $u_{\ast}/u_{{\ast}{\mathrm{t}}}=1.45$ ($1.80$). The dashed line in the upper inset corresponds to terrestrial dunes obtained with $u_{\ast}/u_{{\ast}{\mathrm{t}}} = 1.45$. In the lower inset, we see $L/L_{\mathrm{min}}$ vs $H/H_{\mathrm{min}}$ from calculations of the Arkhangelsky (triangles) and terrestrial dunes (dashed line).} 
     \label{fig:Mars_fields}
 \end{center}
 \end{figure}

Successful modeling of the formation of dunes was recently achieved \cite{Sauermann_et_al_2001,Kroy_et_al_2002}. The dune model consists of a system of continuum equations in two space dimensions which reproduce the shape of dunes, the wind profile and the sand flux and provide excellent quantitative agreement with measurements \cite{Sauermann_et_al_2003}. The sand bed can exchange grains with the moving saltation layer. The local change in the flux, ${\vec{\nabla}}{\cdot}{\vec{q}}$, is proportional to the erosion rate, $\Gamma(x,y)$, which is the difference between the vertical flux of ejected grains and the vertical flux $\phi$ of grains impacting onto the bed: $\Gamma = {\phi}(n - 1)$, where $n$ is the average number of splashed grains. At saturation (${\tau}_{\mathrm{a}} \approx {\tau}_{\mathrm{t}}$), $n$ can be written as \cite{Sauermann_et_al_2001}:
\begin{equation}
n = 1 + {\tilde{\gamma}}{\left({{\frac{{\tau}_{\mathrm{a}}}{{\tau}_{\mathrm{t}}}}-1}\right)} \label{eq:gamma},
\end{equation} 
where ${\tilde{\gamma}}$, the {\em{entrainment rate}} of grains into saltation, determines how fast the system reaches saturation \cite{Sauermann_et_al_2001}. On the other hand, $\phi \equiv |{\vec{q}}|/{\ell}$, where $\ell$ is the average saltation length. The balance $\Gamma(x,y) = \vec{\nabla}{\cdot}{\vec{q}}$ yields a differential equation for $\vec{q}$:
\begin{equation}
{\vec{\nabla}}{\cdot}{\vec{q}}  = {\frac{1}{{\ell}_{\mathrm{s}}}}|{\vec{q}}|{\left({1 - {\frac{|{\vec{q}}|}{q_{\mathrm{s}}}}}\right)}; \ \ {\ell}_{\mathrm{s}} = {\frac{\ell}{\tilde{\gamma}}}{\left[{\frac{{\tau}_{\mathrm{t}}}{{{{{\tau}} - {\tau}_{{\mathrm{t}}}}}}}\right]}, \label{eq:sand_flux}
\end{equation}
where $q_{\mathrm{s}} = |\vec{q}|(\tau - {\tau}_{\mathrm{t}})/(\tau - {\tau}_{\mathrm{a}})$ is the saturated flux \cite{Sauermann_et_al_2001}, and ${\ell} = v_z^{\mathrm{eje}}(2v_{\mathrm{g}}/g)$, where $v_z^{\mathrm{eje}}$ is the initial vertical velocity of the grains, $v_{\mathrm{g}}$ their mean velocity and $g$ is gravity. Further, $v_z^{\mathrm{eje}} = {\alpha}{{\Delta}v_{\mathrm{hor}}}$, where ${\Delta}v_{\mathrm{hor}}$ is the horizontal velocity gain of the particle after one saltation trajectory, and $\alpha$ is an effective restitution coefficient for the grain-bed interaction \cite{Sauermann_et_al_2001}. In this manner, ${\ell} = (1/r)[2{v_{\mathrm{g}}^2}{\alpha}/g]$, with $r \equiv {v_{\mathrm{g}}}/{\Delta}v_{\mathrm{hor}}$, and the saturation length, ${\ell}_{\mathrm{s}}$, may be then written as:
\begin{equation}
{\ell}_{\mathrm{s}} = {\frac{1}{\gamma}}{\left[{\frac{2{v_{\mathrm{g}}^2}{\alpha}/g}{{\left({{u_{\ast}}/u_{{\ast}{\mathrm{t}}}}\right)}^2 - 1}}\right]}, \label{eq:saturation_length}
\end{equation}
where $\gamma \equiv r{\tilde{\gamma}}$. 

The dune model can be sketched as follows: (i) the wind shear stress over the topography is calculated using the algorithm of Weng {\em{et al.}} (1991) \cite{Weng_et_al_1991}. The wind velocity $u$ over a flat ground increases logarithmically with the height $z$: $u(z) = u_{\ast}{\kappa}^{-1}{\ln{z/z_0}}$, where $\kappa= 0.4$ is the von K\'arm\'an constant and $z_0$ is the aerodynamic roughness. A dune introduces a perturbation in the shear stress whose Fourier-transformed components are
\begin{equation}
{\tilde{{\hat{\tau}}}}_x\!=\!{\frac{2\,h(k_x,k_y){k}_{x}^2}{|k|\,U^2(l)}}\!{\left[\!{1\!+\!\frac{2{\ln({\cal{L}}|k_x|)\!+\!4{\epsilon}\!+\!1\!+\!{\mbox{i}}\,{\mbox{sign}}(k_x){\pi}}}{\ln{\left({l/z_0}\right)}}}\!\right]} \label{eq:tau_x}
\end{equation}
and
\begin{equation}
{\tilde{{\hat{\tau}}}}_y={\frac{2\,h(k_x,k_y)k_xk_y}{|k|\,U^2(l)}}, \label{eq:tau_y}
\end{equation}
where the coordinate axes $x$ and $y$ are parallel, respectively, perpendicular to the wind direction, $k_x$ and $k_y$ are wave numbers, $|k|\!=\!\sqrt{k_{x}^2+k_{y}^2}$ and $\epsilon\!=\!0.577216$ (Euler's constant). ${\cal{L}}$ is the horizontal distance between the position of maximum height, $H_{\mathrm{max}}$, and the position of the windward side where the height is $H_{\mathrm{max}}/2$ \cite{Weng_et_al_1991}. $U(l)\!=\!u(l)/u(h_{\mathrm{m}})$ is the undisturbed wind velocity at height $l\!=\!{2{\kappa}^2{\cal{L}}}/{\ln{{l}/z_0}}$ normalized by the velocity at the reference height $h_{\mathrm{m}}\!=\!{\cal{L}}/{{\sqrt{{\log{{\cal{L}}/z_0}}}}}$, which separates the middle and upper flow layers \cite{Weng_et_al_1991}. The shear stress in the direction $i$ ($i\!=\!x,y$) is then given by ${\vec{{\tau}}}_i\!=\!{\hat{{i}}}\left[{{\tau}_0{(1+{{\hat{\tau}}}_i)}}\right]$, where ${\tau}_0$ is the undisturbed shear stress; (ii) next, the sand flux $\vec{q}$ ($x,y$) is calculated using eqs. (\ref{eq:sand_flux}) and ({\ref{eq:saturation_length}}); (iii) the change in surface height $h(x,y)$ is computed from mass conservation: ${{\partial}h}/{{\partial}t} \!= \!- {\vec{\nabla}}{\cdot}{\vec{q}}/{{\rho}_{\mathrm{sand}}}$, where ${\rho}_{\mathrm{sand}} \!= \!0.62 {\rho}_{\mathrm{grain}}$ is the bulk density of the bed \cite{Sauermann_et_al_2001}; and (iv) if sand deposition leads to slopes that locally exceed the angle of repose, the unstable surface relaxes through avalanches in the direction of the steepest descent, and the separation streamlines are introduced at the dune lee \cite{Kroy_et_al_2002}. Each streamline is fitted by a third order polynomial connecting the brink with the ground at the reattachment point \cite{Kroy_et_al_2002}, and defining the ``separation bubble'', in which the wind and the flux are set to zero. The model is evaluated by performing steps i) through iv) computationally in a cyclic manner.

In this letter, our aim is to reproduce the shape of Mars dunes using the martian present atmospheric conditions. Most of the relevant model parameters are known for Mars: The pressure, $P$, and temperature, $T$, which have been measured in several places on Mars \cite{MGSRS}, determine ${\rho}_{\mathrm{fluid}}$. Also known are the average grain diameter $d$ and density, ${\rho}_{\mathrm{grain}} = 3200$ kg$/$m$^3$, and gravity, $g=3.71$ m$/$s$^2$, while $u_{{\ast}{\mathrm{t}}}$ is calculated as in ref. \cite{Iversen_and_White_1982}. 

One very common type of dune on Mars are {\em{barchans}} \cite{Bourke_et_al_2004}. They have one slip face and two horns, and propagate on bedrock under conditions of uni-directional wind \cite{Bagnold_1941}. Indeed, to develop a slip face, they must reach a minimum size. In the Arkhangelsky Crater on Mars (fig. 1), the minimal dune width is $W_{\mathrm{min}} \approx 200$ m and the corresponding length $L_{\mathrm{min}} \approx 400$ m. On the other hand, $W_{\mathrm{min}}$ is around 13 times the saturation length, ${\ell}_{\mathrm{s}}$ \cite{Parteli_et_al_2006}. In this manner, $W_{\mathrm{min}}$ yields, through eq. (\ref{eq:saturation_length}), the wind strength $u_{\ast}/u_{{\ast}{\mathrm{t}}}$ on Mars.

However, there is {\em{one}} unknown quantity for Mars which we need in order to solve the sand transport equations: $\gamma$, which appears in eq. (\ref{eq:saturation_length}). Sauermann {\em{et al.}} (2001) \cite{Sauermann_et_al_2001} obtained $\gamma = 0.2$ for saltation on Earth \cite{Sauermann_et_al_2001} by comparing with direct measurements of saturation transients \cite{McEwan_and_Willetts_1991,Butterfield_1993}. Such measurements are not available for Mars.

We perform our calculations using open boundaries and a constant upwind $u_{\ast}$ in $x$ direction at the inlet starting with a Gaussian hill, which evolves until displaying the linear relations between height $H$, width $W$ and length $L$ of barchans \cite{Sauermann_et_al_2001,Parteli_et_al_2006}.

For a constant $\gamma$, there is {\em{one}} value of $u_{\ast}$ which reproduces the minimal dune width $W_{\mathrm{min}} = 200$ m in the Arkhangelsky Crater, where $P = 5.5$ mb and $T = 210$ K \cite{MGSRS}, which gives $u_{{\ast}{\mathrm{t}}} \approx 2.12$ m$/$s. We obtained a surprising result: If we take the same $\gamma$ as on Earth, the value of $u_{\ast}$ which gives $W_{\mathrm{min}} = 200$ m is around $6.0$ m$/$s. This value is far too large to be realistic. The largest peaks of martian $u_{\ast}$ range between $2.2$ and $4.0$ m$/$s \cite{Moore_1985,Sullivan_et_al_2005}. To obtain the correct $W_{\mathrm{min}}$ using $u_{\ast}$ within this maximum range, we must take ${\gamma}$ at least one order of magnitude larger than on Earth. We must find how much the martian ${\tilde{\gamma}}\!=\!{\mbox{d}}n/{{\mbox{d}}({\tau}_{\mathrm{a}}/{\tau}_{\mathrm{t}})}$ (eq. (\ref{eq:gamma})) differs from the one on Earth. 

Anderson and Haff (1988) \cite{Anderson_and_Haff_1988} showed that the number of splashed grains is proportional to the velocity $v_{\mathrm{imp}}$ of the impacting grains. Let us rescale $v_{\mathrm{imp}}$ with $v_{\mathrm{eje}} = {\sqrt{gd}}$, which is the velocity necessary to escape from the sand bed \cite{Andreotti_2004}. Further, $v_{\mathrm{imp}}$ scales with the mean grain velocity $v_{\mathrm{g}}$ \cite{Sauermann_et_al_2001}. In this manner, we obtain ${\tilde{\gamma}} \propto v_{\mathrm{g}}/{\sqrt{gd}}$. 

Moreover, $v_{\mathrm{g}}$ scales with $u_{{\ast}{\mathrm{t}}}$ \cite{Sauermann_et_al_2001} and has only a very weak dependence on $u_{\ast}$ which we neglect, and thus ${\tilde{\gamma}} \propto u_{{\ast}{\mathrm{t}}}/{\sqrt{gd}}$. Since we know that $\gamma = 0.2$ on Earth, where $g = 9.81$ m$/$s$^2$, $d = 250$ $\mu$m and $u_{{\ast}{\mathrm{t}}} = 0.218$ m$/$s \cite{Sauermann_et_al_2001}, we obtain 
\begin{equation}
\gamma = 0.045 \frac{u_{{\ast}{\mathrm{t}}}}{\sqrt{gd}}. \label{eq:equation_for_gamma}
\end{equation}
Substituting eq.~(\ref{eq:equation_for_gamma}) into eq.~(\ref{eq:saturation_length}), we obtain a closed set of sand transport equations from which the value of the wind friction speed $u_{\ast}$ on Mars can be determined. 

Eq.~(\ref{eq:equation_for_gamma}) gives $\gamma \approx 2.24$ in the Arkhangelsky Crater, and $W_{\mathrm{min}} = 200$ m is reproduced for $u_{\ast}/u_{{\ast}{\mathrm{t}}} = 1.45$ or $u_{\ast} = 3.07$ m$/$s, which gives ${\ell}_{\mathrm{s}} \approx 15.5$ m. Fig. 1 shows that $u_{\ast} = 3.07$ m$/$s reproduces not only the minimal dune but also the dependence of the dune shape on its size. Furthermore, we tested the scaling relation (\ref{eq:equation_for_gamma}) with a second martian barchan field which is near the north pole, and where $W_{\mathrm{min}} \approx 80$ m. At the location of the field, $P=8.0$ mb and $T=190$ K \cite{MGSRS}, and thus $u_{{\ast}{\mathrm{t}}} \approx 1.62$ m$/$s. From $W_{\mathrm{min}}$, we obtain ${\ell}_{\mathrm{s}} \approx 6.0$ m, which gives $u_{\ast}/u_{{\ast}{\mathrm{t}}} \approx 1.8$ or $u_{\ast} = 2.92$ m$/$s. The plot in fig. 1 shows that the behaviour $L$ against $W$ of the barchans in both studied fields is well captured by the model. 

It is interesting that the $u_{\ast}$ obtained for the north polar field is very similar to that in the Arkhangelsky Crater, although $u_{{\ast}{\mathrm{t}}}$ is lower in the north polar field due to the higher ${\rho}_{\mathrm{fluid}}$ (table 1). 

Summarizing, we discovered that the rate $\gamma$ at which grains enter saltation on Mars is about one order of magnitude higher than on earth. Taking this finding into account, we found that the wind velocity which reproduces the size and shape of martian barchans is $u_{\ast} \approx 3.0 \pm 0.1$ m$/$s, which is well within the range of estimated wind speeds on Mars \cite{Moore_1985,Sullivan_et_al_2005}. We found that Mars grains travel with a velocity $v_{\mathrm{g}}$ ten times higher than that of terrestrial grains, as observed in wind tunnel experiments for Mars \cite{White_1979}. Such high-speed grains produce much larger splash events than on Earth \cite{Marshall_et_al_1998}, and lead to higher $\gamma$ values (table 1).

\begin{table}
\begin{center} 
\begin{tabular}{|c|c|c|c|c|}
\hline
Barchan field & ${\rho}_{\mathrm{fluid}}$ (kg$/$m$^3$) & $u_{{\ast}{\mathrm{t}}}$ (m$/$s) & $v_{\mathrm{g}}$ (m$/$s) & $\gamma$  \\ \hline \hline
Arkhangelsky & $0.014$ & $2.12$ & $17.8$ & $2.24$ \\ \hline
$77.6^{\circ}$N, $103.6^{\circ}$W & $0.022$ & $1.62$ & $12.3$ & $1.71$ \\ \hline
Earth & $1.225$ & $0.22$ & $1.5$ & $0.20$  \\ \hline 
\end{tabular}
\end{center}
\vspace{-0.5cm}
\caption{Main quantities controlling saltation on Mars and on Earth. $u_{{\ast}{\mathrm{t}}}$ is around $80\%$ of the direct entrainment threshold $u_{{\ast}{\mathrm{ft}}}=A\,\sqrt{{\rho}_{\mathrm{grain}}gd/{\rho}_{\mathrm{fluid}}}$, where $A = 0.129 {\left[{{{\left({1 + 6.0 \times 10^{-7}/{{{\rho}_{\mathrm{grain}}}gd^{2.5}}}\right)}^{0.5}}/{{\left({1.928{\mbox{Re}}_{{\ast}{\mathrm{ft}}}^{0.092}}-1\right)}^{0.5}}}\right]}$ \cite{Iversen_and_White_1982}, ${\mbox{Re}}_{{\ast}{\mathrm{ft}}} \equiv u_{{\ast}{\mathrm{ft}}}d{\rho}_{\mathrm{fluid}}/{\eta}$, and $\eta$ is the viscosity of CO$_2$ gas.} \label{tab:parameters_fields}
\end{table}

Our results show that the larger splash events have a crucial implication for the formation of dunes on Mars. While on one hand the lower martian $g$ and ${\rho}_{\mathrm{fluid}}$ result in longer grain trajectories \cite{White_1979}, the higher rate at which grains enter saltation on Mars shortens the saturation transient of the flux, which determines the minimal dune size. The distance of flux saturation, ${\lambda}_{\mathrm{s}}$, is around six times the characteristic length ${\ell}_{\mathrm{s}}$ \cite{Sauermann_et_al_2001}, which effectively increases with $\ell$. Here we found that ${\lambda}_{\mathrm{s}}$ on Mars is indeed reduced by the larger martian splash. This explains the previously reported \cite{Kroy_et_al_2005} failure of the scaling $W_{\mathrm{min}} \propto {\ell}$ for Mars. 

\begin{figure}
  \begin{center}
   \includegraphics[width=0.75 \columnwidth]{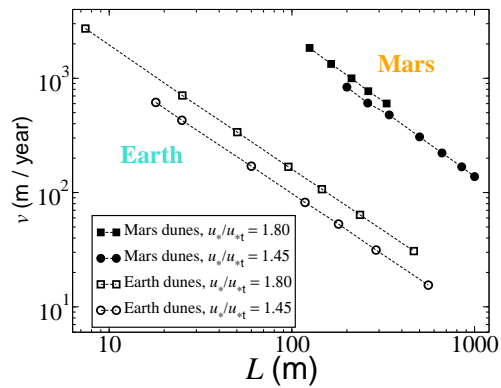}	
 \caption{Dune velocity $v$ as function of dune length $L$. We see that Mars dunes (filled symbols) move typically ten times faster than Earth dunes (empty symbols) of same $L$, obtained with similar values of $u_{\ast}/u_{{\ast}{\mathrm{t}}}$ as on Mars.}
     \label{fig:dune_velocity}
 \end{center}
 \end{figure}

The values of $u_{\ast}/u_{{\ast}{\mathrm{t}}}$ obtained for Mars are within the range of the ones measured in terrestrial barchan fields \cite{Sauermann_et_al_2003}. Indeed, we see in fig. \ref{fig:dune_velocity} that, for the same value of $u_{\ast}/u_{{\ast}{\mathrm{t}}}$, Mars barchans would move ten times faster than those on Earth. 

Many exotic and up to now unexplained dune forms have also been observed on Mars. Dunes as those in fig. 3 cannot be obtained from calculations with a uni-directional wind. We found that the dune shapes in fig. 3 may be obtained with a {\em{bimodal}} wind regime. In our calculations, the wind {\em{alternates}} its direction periodically with frequency $1/T_{\mathrm{w}}$ forming an angle $2|{\theta}_{\mathrm{w}}|$ as sketched in fig. 3a$^{\prime}$. In both directions the strength is the same, namely $u_{\ast} = 3.0$ m$/$s. In this manner, the value of $u_{\ast}/u_{{\ast}{\mathrm{t}}}$ is particular to each field, since $u_{{\ast}{\mathrm{t}}}$ depends on the field location (table 2). To simulate the change of wind direction, we rotate the field by an angle $2{\theta}_{\mathrm{w}}$, keeping the wind direction constant. The separation bubble, thus, adapts to the wind direction after rotation of the field. We use open boundaries as in the calculations of barchan dunes. Initial condition is a gaussian hill as before, whose volume is taken according to the volume of the dune. 

The angle $2{\theta}_{\mathrm{w}}$ between the wind directions determines which of the different forms in fig. 3 is obtained. We found that a barchan moving in the resulting wind direction is always obtained if $|{\theta}_{\mathrm{w}}| < 50^{\circ}$. If $|{\theta}_{\mathrm{w}}| \approx 50^{\circ}$, the dune shape in fig. 3a$^{\prime}$ is achieved. And for $|{\theta}_{\mathrm{w}}|$ between $50^{\circ}$ and $70^{\circ}$, elongated dune forms as those in fig. 3b$^{\prime}$ are obtained, which elongate further and further in time. As $|{\theta}_{\mathrm{w}}| \longrightarrow 90^{\circ}$, a barchanoidal form of alternating slip face position appears.

\begin{figure}
  \begin{center}
   \includegraphics[width=1.0 \columnwidth]{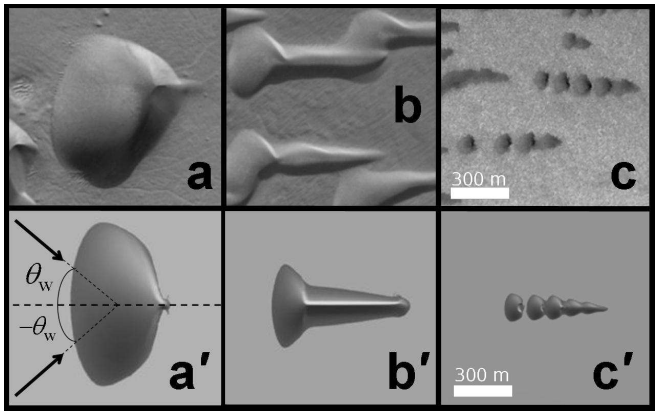}
 \caption{MOC images on top (courtesy of NASA/JPL/MSSS) and calculations obtained with bimodal wind regimes on bottom. A sketch showing the definition of the angle ${\theta}_{\mathrm{w}}$ of the wind direction (arrows) is shown in a$^{\prime}$. The wind changes its direction with frequency $1/T_{\mathrm{w}}$. We chose $T_{\mathrm{w}} =$ $2.9$ days, $5.8$ days and $0.7$ day to obtain dunes in a$^{\prime}$, b$^{\prime}$ and $c^{\prime}$, respectively. The dune in a$^{\prime}$ has been obtained with $|{\theta}_{\mathrm{w}}| = 50^{\circ}$ and the dune in b$^{\prime}$ with $|{\theta}_{\mathrm{w}}|=70^{\circ}$. In c$^{\prime}$, a linear dune obtained with $|{\theta}_{\mathrm{w}}| = 60^{\circ}$ decays into barchans after the angle $|{\theta}_{\mathrm{w}}|$ is reduced to $40^{\circ}$. Dune in b$^{\prime}$ elongates to the right further and further with time, which is not observed for dune in a$^{\prime}$. Dune in c$^{\prime}$ decays further until only a string of rounded barchans is visible. }
     \label{fig:longdunes}
 \end{center}
 \end{figure}
\begin{table}
\begin{center} 
\begin{tabular}{|c|c|c|c|c|}
\hline
Field & location & ${\rho}_{\mathrm{fluid}}$ (kg$/$m$^3$) & $u_{{\ast}{\mathrm{t}}}$ (m$/$s) \\ \hline \hline
fig. 3a & $48.6^{\circ}$S, $25.5^{\circ}$W & $0.017$ & $1.89$ \\ \hline 
fig. 3b & $49.6^{\circ}$S, $352.9^{\circ}$W & $0.014$ & $2.06$ \\ \hline 
fig. 3c & $76.4^{\circ}$N, $272.9^{\circ}$W & $0.03$ & $1.35$ \\ \hline 
\end{tabular}
\end{center}
\vspace{-0.5cm}
\caption{For each dune field in fig. 3, the fluid density ${\rho}_{\mathrm{fluid}}$ and the threshold $u_{{\ast}{\mathrm{t}}}$ is calculated from the local pressure and temperature which are taken from ref. \cite{MGSRS}. In spite of the broad range of $u_{{\ast}{\mathrm{t}}}$, all dune forms in fig. 3 have been obtained with one single value of $u_{\ast} = 3.0$ m$/$s.} \label{tab:parameters_longdunes}
\end{table}

Moreover, we found that the structure observed in the dune field of fig. 3c can be obtained by a change in the local wind regime. The dune shape in fig. 3c$^{\prime}$ has been obtained in the following manner: (i) first, an elongated dune form as the one in fig. 3b$^{\prime}$ is formed with an angle $|{\theta}_{\mathrm{w}}| = 60^{\circ}$; (ii) next, the angle $|{\theta}_{\mathrm{w}}|$ has been suddenly reduced to $40^{\circ}$. Thereafter, the linear dune becomes unstable and decays into a string of rounded barchans seen in fig. 3c. 

Each one of the dune shapes in fig. 3 can only be achieved if the time $T_{\mathrm{w}}$ is of the order of $1-5$ days. If the period is too large, of the order of $1$ month, then the dune evolves into a barchanoidal form.

In conclusion, we found from calculations of martian barchans using the present atmospheric conditions of Mars, that the splash on Mars must be about ten times larger than on Earth. Furthermore, we obtained a general equation for the entrainment rate of grains into saltation which could be used in future calculations of other dune forms on Mars, Venus or Titan. We also found that winds on Mars don't exceed $3.0 \pm 0.1$ m$/$s, which can explain the shape of elongated dunes formed by a bimodal wind. It would be interesting to make a full microscopic simulation for the saltation mechanism of Mars similar to the one that was recently performed by Almeida {\em{et al.}} (2006) \cite{Almeida_et_al_2006} to confirm our findings microscopically.

\acknowledgments
We acknowledge Kenneth Edgett, Keld Rasmussen, Bruno Andreotti and Orencio Dur\'an for discussions. This research was supported in part by The Max-Planck Prize and the Volkswagenstiftung. E. J. R. Parteli acknowledges support from CAPES - Bras\'{\i}lia/Brazil.

\end{document}